# A NEW ANALYTIC STATISTICAL MECHANICAL MODEL FOR POLAR FLUIDS

**Chih-An HWANG[(*)] and G.Ali MANSOORI[(**)]**

University of Illinois at Chicago (M/C 063) Chicago, Illinois 60607-7052, U.S.A.

## ABSTRACT

A new analytic statistical mechanical model for polar fluids with an intermolecular pair potential consisting of a hard core and multiple attractive Yukawa tails has been developed. It includes all the leading terms in the orientationally averaged electrostatic energies up to quadrupole-quadrupole interactions. This model is capable of producing analytic expressions for all the thermodynamic properties of both non-polar and polar fluids. We have evaluated the Yukawa potential parameters to represent the dispersion potential from several methods and formulated other Yukawa potentials to represent electrostatic potentials in simple closed forms, which then could be calculated directly. Therefore, introducing an arbitrary number of Yukawa potentials does not introduce any further computational difficulty. Thermodynamic properties of three representative fluids (argon, carbon dioxide, and methyl chloride) are predicted with this model. The results indicate good agreement between experiment and the predictions from this model, given only the Yukawa intermolecular potential parameters and multipole moments.

_______________________________________

(*).   Present address: Chevron Phillips Chemical Co., Humble, Texas, USA
        Email: _hwangca88@gmail.com_
(**). Email: _mansoori@uic.edu_

## INTRODUCTION

Recent developments in statistical thermodynamics indicate that fundamental statistical mechanical theories are capable of producing simple and accurate analytic relations for thermodynamic properties. The discovery of new approximations for the radial distribution function (RDF) has led to significant progress in classical statistical mechanical study of dense fluids (Hill, 1956; Rice and Gray, 1965;Egelstaff, 1967). There are various closures to the Ornstein-Zernike (OZ) integral equation for the direct correlation function (Ornstein and Zernike, 1914). The "closure" supplies the needed relation between the direct correlation function and the RDF. Among them, the Percus-Yevick (1958) closure which relates the direct correlation function to the RDF and intermolecular potential is rather simple and the most satisfactory one. In addition, the popularity of PY closure has been achieved by its closed form solution for the hard-sphere potential (Wertheim, 1963; Thiele, 1963). An alternate approach is the Yukawa closure of the OZ equation under the mean spherical approximation (MSA) first adopted by Waisman (1973). H∅ye and Stell (1977) then examined the thermodynamic properties of the MSA for simple fluids. The later approach together with thermodynamic consistency requirements was used in the model developed by Mansoori and Kioussis (1985) to produce simple analytic expressions for all thermodynamic properties of the fluid system. This model for the hard-core fluid with a Yukawa tail has been compared with exact (Monte Carlo) results with success (Mansoori and Kioussis, 1985). It is of interest to apply this model to real fluids.

The goal of this work has been twofold: 1) to evaluate the Yukawa potential parameters to represent the dispersion potential from several methods, and 2) to formulate other Yukawa potentials to represent each inverse power of separation distance in electrostatic potentials, which include all the leading terms in the orientationally averaged electrostatic energy for neutral molecules of the polar fluids. It should be noted that all the leading terms in the orientationally averaged electrostatic energy are of attractive nature. The parameters of the Yukawa potentials representing these attractive potentials have been reduced to simple closed forms and can be calculated directly. Therefore, introducing an arbitrary number of Yukawas will not introduce any further computational difficulty. The resulting hard-core Yukawa-tail model is accurate enough to take the place of the existing empirical approaches of equations of state and has been applied to fluids such as carbon dioxide and methyl chloride with multiple Yukawas. It allows the prediction of thermodynamic properties of fluids of industrial importance.

## DERIVATION OF THE MODEL

Consider a fluid with molecules which possess a pairwise additive intermolecular potential consisting of a hard core and multiple Yukawa tails of the form

$$\phi(x) = \begin{cases} \infty, & x = \dfrac{r}{\sigma} \le 1 \\[2mm] -\sum_i \dfrac{\varepsilon_i}{x} \exp[-z_i(x-1)], & x > 1 \end{cases} \tag{1}$$

where r is the intermolecular separation distance, $\sigma$ is the diameter of the hard core of the molecules, $\varepsilon$ is the parameter describing the depth of attractive well in the intermolecular interaction, and z is an inverse range parameter. The dimensionless residual internal energy, $U^r/RT$, for spherically symmetric molecules is

$$\frac{U^r}{RT} \equiv \beta \Delta u = 2\pi\beta\rho\sigma^3 \int_0^\infty g(x)\phi(x)x^2 dx, \tag{2}$$

where $\beta = (kT)^{-1}$, k is Boltzmann's constant, T is the thermodynamic temperature, R is the universal gas constant, $\Delta u$ is the residual internal energy per particle, $\rho$ is the number density, and g(x) is the radial distribution function of the fluid system. The term "residual" denotes the differences between the properties of a real fluids and the perfect gas at the same temperature and pressure (or density).

H$\varnothing$ye and Stell (1977) have shown that within the mean spherical approximation (MSA), the energy equation of state of a fluid with a hard core can be calculated from

$$\frac{P}{\rho kT} = \frac{P_0}{\rho kT} + \frac{\pi}{3}\rho\sigma^3[g^2(1^+) - g_0^2(1^+)] - \frac{2\pi\rho\sigma^3}{3kT}\int_1^\infty g(x)\frac{d\phi(x)}{dx}x^3 dx, \tag{3}$$

where $P_0$ and $g_o(1^+)$ are the pressure and the contact value of the radial distribution function of the hard-sphere reference system, respectively. Properties with subscript 0 denote the hard-sphere quantities. This result permits the calculation of the MSA energy equation of state with no more difficulty than that required from the pressure equation since the last term in Eq. (3) is exactly the same as that appearing in the pressure equation.

By substituting Eq. (1) into Eqs. (2) and (3) we obtain

$$\Delta u = 2\pi\rho\sigma^3 \int_1^\infty g(x)\phi(x)x^2 dx = -2\pi\rho\sum_i \varepsilon_i e^{z_i}\int_1^\infty xg(x)e^{-z_i x}dx = -2\pi\rho\sum_i \varepsilon_i e^{z_i}\hat{g}(z_i) \tag{4}$$

and the last term in Eq. (3) becomes,

$$-\frac{2\pi\rho\sigma^3}{3kT}\int_1^\infty g(x)\frac{d\phi(x)}{dx}x^3 dx = \frac{2\pi\rho}{3kT}\sum_i \varepsilon_i e^{z_i}\int_1^\infty xg(x)(-z_i x-1)e^{-z_i x}dx$$

$$= \frac{2\pi\rho}{3kT}\sum_i \varepsilon_i e^{z_i}[z_i \hat{g}'(z_i)-\hat{g}(z_i)]$$

therefore

$$\frac{P}{\rho kT} = \frac{P_0}{\rho kT} + \frac{\pi}{3}\rho^*[g^2(1^+) - g_0^2(1^+)] + \frac{2\pi\rho}{3kT}\sum_i \varepsilon_i e^{z_i}[z_i \hat{g}'(z_i)-\hat{g}(z_i)] \qquad (5)$$

in which we have defined the dimensionless quantity $\rho* = \rho\sigma^3$.   The integral in Eq. (4) is the Laplace transform of $xg(x)$ for $s = z$, and for the hard-sphere reference system we use the Carnahan and Starling (1969) expressions,

$$\frac{P_0}{\rho kT} = \frac{1+\theta+\theta^2-\theta^3}{(1-\theta)^3} \qquad (6)$$

for the equation of state, and

$$g_0(1^+; \theta) = \frac{2-\theta}{2(1-\theta)^3} \qquad (7)$$

for the contact value of the radial distribution function, where the packing fraction $\theta = \pi\rho\sigma^3/6 = \pi\rho*/6$.

At this stage the exact RDF, $g(x)$, is approximated with the Percus-Yevick RDF, $g_o(x)$, for hard-spheres. The Laplace transform, $g(z)$ With Hat, of $g_o(x)$ is then given by Wertheim's (1963) analytic solution (for the sake of simplicity the subscript $i$ in the multiple Yukawa expression $z_i$ is omitted),

$$\hat{g}_0(z) = zf_1(\eta;z), \qquad (8)$$

where

This paper appeared in:

$$f_1(\eta;z) = \frac{L(\eta;z)}{12\eta L(\eta;z)+S(\eta;z)e^z} \qquad (9a)$$

and $\eta = \pi \rho d^3/6$. The reduced density $\eta$ and the corresponding hard-sphere

$$S(\eta;z) = (1-\eta)^2 z^3 + 6\eta(1-\eta)z^2 + 18\eta^2 z - 12\eta(1+2\eta), \qquad (9c)$$

diameter d will be considered as "state-dependent effective hard-particle" quantities. They are expected to be dependent on both the density and temperature, and must be determined self-consistently which will be explained in the determination of $\eta(\rho$, T) section. As a result of the above approximation for the RDF, the contact value $g(1^+)$ in Eq.(5) will then be equal to $g_0(1^+;\eta)$ given by Eq.(7), but with $\theta$ replaced by $\eta$.

Upon substituting Eqs.(6) and (8) into (4) and (5), we find the expressions

$$\Delta u/\rho = F(\eta;\sigma^3;\varepsilon_i;z_i) = -2\pi\sigma^3 \sum_i \varepsilon_i z_i e^{z_i} f_1(\eta;z_i) \qquad (10)$$

$$\frac{P}{\rho kT} = \frac{1+\theta+\theta^2-\theta^3}{(1-\theta)^3} + \frac{\pi}{3}\rho [g_0^2(1^+;\eta) - g_0^2(1^+;\theta)] + \frac{2\pi\rho}{3kT}\sum_i \varepsilon_i z_i^2 e^{z_i} f_2(\eta;z_i), \qquad (11)$$

Where

$$f_2(\eta;z) = \partial f_1(\eta;z)/\partial z$$
$$= e^z \{L_1(\eta;z)S(\eta;z) - L(\eta;z)[S(\eta;z)+S_1(\eta;z)]\}/\{12\eta L(\eta;z)+S(\eta;z)e^z\}^2, \qquad (12a)$$

$$L_1(\eta;z) = \partial L(\eta;z)/\partial z = 1 + \eta/2, \qquad (12b)$$

and

$$S_1(\eta;z) = \partial S(\eta;z)/\partial z = 3(1-\eta)^2 z^2 + 12\eta(1-\eta)z + 18\eta^2 \qquad (12c)$$

Here again the subscript i in $z_i$ has been omitted in Eq. (12) for simplicity. Equations (10)-(12) constitute the basic equations for determining the internal energy and compressibility factor for the hard-core Yukawa-tail fluid, provided that the reduced density $\eta$ is known as a function of density and temperature. In

This paper appeared in:

$$B(T) = \frac{2\pi}{3}\sigma^3 [1 - \frac{1}{kT}\sum_i \varepsilon_i (\frac{z_i^2+3z_i+3}{z_i^2})] \qquad (13)$$

addition, the expression for the second virial coefficient, as derived from Eq. (11), is

The quantity $(2/3)^1 \sigma^3$ is the hard-sphere second virial coefficient. Note that as $\varepsilon$ approaches zero, Eq. (13) reduces to the hard-sphere result. It reduces to the sticky hard-sphere second virial coefficient when z goes toward infinity.

## EXTENSION TO POLAR FLUIDS

In order to extend this model for polar fluids, we formulate Yukawa potentials to represent the electrostatic attractive potentials. Recall that the angle-averaged, pair-interaction, intermolecular potential-energy function has the form (Massih and Mansoori, 1983)

$$\overline{U}(r,T) = 4\varepsilon_{LJ}[(\frac{\sigma_{LJ}}{r})^{12} - (\frac{\sigma_{LJ}}{r})^{6}]$$
$$- (\frac{\mu_i^2 \mu_j^2}{3kT} + \alpha_i \mu_j^2 + \alpha_j \mu_i^2) \frac{1}{r^6} - (\frac{\mu_i^2 Q_j^2 + \mu_j^2 Q_i^2}{2kT}) \frac{1}{r^8} - (\frac{7 Q_i^2 Q_j^2}{5 \ kT}) \frac{1}{r^{10}} \qquad (14)$$

or

$$\overline{U}(r,T) = U_{disp}(r) - \frac{\alpha_6(T)}{r^6} - \frac{\alpha_8(T)}{r^8} - \frac{\alpha_{10}(T)}{r^{10}} \qquad (15)$$

where $\alpha_i$, $\mu_i$, and $Q_i$ are the polarizability, dipole moment, and quadrupole moment of the molecule i, respectively. In Eq. (14) the Lennard-Jones (LJ) potential is used to represent the dispersion potential and all the significant leading terms up to quadrupole-quadrupole interactions in the orientationally averaged electrostatic energy are included. These permanent multipole moments of neutral polar molecules are about the same order of magnitude as dispersion potential which is of the order of $10^{-14}$ erg. The relative value is dispersion potential > permanent multipole moments > induced moments. We plan to use one extra Yukawa, $\varepsilon_i \ e^{-zi}$

$$\frac{\alpha_{10}}{r^{10}} + \frac{\alpha_8}{r^8} = \frac{\varepsilon_1}{x} e^{-z_1(x-1)} + \frac{\varepsilon_2}{x} e^{-z_2(x-1)} \qquad (16)$$

$$\varepsilon_i = \alpha_{n-2(i-1)}(T) / \sigma^{n-2(i-1)}$$

$(x-1)$ /x, to represent one inverse power of separation distance $\alpha_n$ /$r^n$ term. The procedure for determining the Yukawa potential parameters to represent these electrostatic potentials is as follows. First, for the highest n (=10) the equality $\varepsilon_1$ $e^{-z_1(x-1)}$ /x = $\alpha_n$ /$r^n$ is to be satisfied with the boundary conditions at r = σ and r = $r_n$ = $C_1 \bullet \sigma$, where $C_1$ is a constant and $r_n$ represents the maximum intermolecular separation distance in the calculating procedure of $\alpha_n$ /$r^n$ term. Beyond $r_n$ the contribution of $\alpha_n$ /$r^n$ term is small and can be neglected. Next, we include one more Yukawa term for the next higher n (=8) but try to reduce accumulated errors in representing electrostatic potentials; i.e., let where $\varepsilon_1$ and $z_1$ (the highest n) are known, and $\varepsilon_2$ and $z_2$ are to be determined from boundary conditions at r = σ and r = $r_8$ =$C_2 \bullet \sigma$. Similar steps proceed to lower n values. It is obvious that this calculating procedure can be extended to higher n values, say n=12. After mathematical manipulation, one can obtain:

$$z_1 = \ln\left(\frac{1}{C_1^9}\right) / (1 - C_1) \tag{18}$$

$$z_i = \ln\left\{ \sum_{k=0}^{} \frac{\alpha_{n-2k}}{\alpha_{n-2(i-1)}\sigma^{2(i-1)-2k}} \left[ \frac{1}{C_i^{(n-1)-2k}} - e^{z_{1+k}(1-C_i)} \right] \right\} / (1 - C_i) \tag{17}$$

where n represents the highest $\alpha_n$ /$r^n$ term, 1+k < i and i=1, 2,.... For example for carbon dioxide, n=10 and i=1:

The value of $C_i$, which in turn varies the value of $z_i$, should be determined so as to reduce the accumulated errors in representing electrostatic potentials. Again we use carbon dioxide as an example and demonstrate how we determine $\varepsilon_1$ and $z_1$ for representing the quadrupole-quadrupole potential. First, we assume σ = $\sigma_{LJ}$ and a reasonable value of $C_1$, and we define temporary $\varepsilon_1$ and $z_1$ values. We next fit the second virial coefficients to determine the Yukawa parameters σ and $\varepsilon_d$, of which $z_d$ has a known value of 1.8, using Eq. (13) or any thermodynamic properties. The subscript d denotes the Yukawa parameters to represent the dispersion potential. Using these determined parameters ($\varepsilon_1$ has been revised according to Eq. (17)), we can predict other thermodynamic properties such as pressure, Eq. (11), and obtain the optimal value of $C_1$ when minimum deviation from the experimental pressures has been achieved.

## DETERMINATION OF η(ρ, T)

We now explain how one can determine $\eta(\rho,T)$, the "state-dependent effective hard-particle" reduced density, by imposing thermodynamic consistency between the expressions relating the derivatives of the Helmholtz free energy with the internal energy and pressure. In connection with this consistency requirement, the next question that arises is the choice of the functional form of the entropy of the hard-core Yukawa-tail fluid.

In analogy with the model reported by Mansoori and Kioussis (1985) we model the residual entropy for the Yukawa fluid, $\Delta s$, with the residual entropy for hard spheres, which is of the form $\Delta s = \Delta s(\eta_s)$. Here, $\eta_s = (1/6)\rho d_s^3$ is another "state-dependent effective hard-particle" reduced density, analogous to $\eta$, and must also be determined self-consistently as a function of density and temperature. We use the Carnahan and Starling expression for $\Delta s$ which is given by

$$\Delta s = -kG(\eta_s) = -k\frac{\eta_s(4-3\eta_s)}{(1-\eta_s)^2}. \tag{19}$$

From Eqs. (10) and (19) it immediately follows that

$$\frac{A^r}{RT} \equiv \beta\rho\Delta a = \beta\rho F[\eta;\sigma^3(T);\varepsilon_i(T);z_i(T)] + G\{\eta_s[\rho;\sigma^3(T);\beta(T)]\}, \tag{20}$$

where $A^r$ is the residual Helmholtz free energy and $\beta = (kT)^{-1}$.

Thermodynamic consistency requires that the following relations be satisfied,

$$\Delta u = [\partial(A^r/RT)/\partial\beta]_\rho \tag{21}$$

$$P^r = \rho^2[\partial A^r/\partial\rho]_\beta \tag{22}$$

Applying Eqs. (21) and (22) to Eq.(20) we obtain

$$\beta\rho\left\{\left(\frac{\partial F}{\partial\eta}\right)\left(\frac{\partial\eta}{\partial\beta}\right)_{\rho,\sigma^3,\varepsilon_i z_i} + \left(\frac{\partial F}{\partial\sigma^3}\right)\left(\frac{\partial\sigma^3}{\partial\beta}\right)_{\rho,\sigma,z_i\nu_i} + \sum_i\left[\left(\frac{\partial F}{\partial\varepsilon_i}\right)\left(\frac{\partial\varepsilon_i}{\partial\beta}\right)_{\rho,\sigma^3,\varepsilon_{j\neq i}z_i\nu_i} + \left(\frac{\partial F}{\partial z_i}\right)\left(\frac{\partial z_i}{\partial\beta}\right)_{\rho,\sigma^3,\varepsilon_i z_{j\neq i}\nu_i}\right]\right\}$$

$$+ \frac{\partial G}{\partial\eta_s}\left\{\left(\frac{\partial\eta_s}{\partial\beta}\right)_{\rho,\sigma^3} + \left(\frac{\partial\eta_s}{\partial\sigma^3}\right)_{\rho,\beta}\left(\frac{\partial\sigma^3}{\partial\beta}\right)_\rho\right\} = 0 \tag{23}$$

$$\frac{P^r}{\rho^2} = F + \rho(\frac{\partial F}{\partial \eta})(\frac{\partial \eta}{\partial \rho})_\beta + \frac{1}{\beta}(\frac{\partial G}{\partial \eta_s})(\frac{\partial \eta_s}{\partial \rho})_\beta . \tag{24}$$

This completes the thermodynamic consistency requirements; i.e., Eqs. (23) and (24) consist of the basic equations for determining $\eta$ and $\eta_s$.

In the high temperature regime, the structure of the real fluid should be the same as that of the hard-sphere reference system, and hence one expects both $\eta$ and $\eta_s$ to approach the hard-sphere reduced density $\theta$ in the limit. Thus, one can assume $\eta$ and $\eta_s$ have the form

$$\eta(\rho,\beta) = \theta(\rho) + e^{\beta \varepsilon_0 \rho} - 1 \tag{25a}$$

And

$$\eta_s(\rho,\beta) = \theta(\rho) + e^{\beta \varepsilon_{0s} \rho} - 1 \tag{25b}$$

where $\varepsilon_0$ and $\varepsilon_{0s}$ are chosen in above forms only to provide dimensionless quantities in the exponents.

Upon differentiating Eqs. (25a) and (25b), inserting the resultant expressions in the basic Eqs. (23) and (24), and replacing $P^r$ in Eq. (24) by Eq. (11), one obtains

$$\beta\rho\{(\frac{\partial F}{\partial \eta})\frac{1}{\beta}(\eta-\theta+1)\ln(\eta-\theta+1) + (\frac{\partial F}{\partial \sigma^3})(\frac{\partial \sigma^3}{\partial \beta})_{\rho,\varepsilon_i,z_i,\eta} + \sum_i[(\frac{\partial F}{\partial \varepsilon_i})(\frac{\partial \varepsilon_i}{\partial \beta})_{\rho,\sigma^3,\varepsilon_{j\neq i},z_i,\eta}$$

$$+ (\frac{\partial F}{\partial z_i})(\frac{\partial z_i}{\partial \beta})_{\rho,\sigma^3,\varepsilon_{j\neq i},z_i,\eta}]\} + \frac{\partial G}{\partial \eta_s}\{\frac{1}{\beta}(\eta_s-\theta+1)\ln(\eta_s-\theta+1) + (\frac{\partial \eta_s}{\partial \sigma^3})_{\rho,\beta}(\frac{\partial \sigma^3}{\partial \beta})_\rho\} = 0 \tag{26}$$

and

$$\beta\rho\{F + \frac{\partial F}{\partial \eta}[\theta+(\eta-\theta+1)\ln(\eta-\theta+1)]\} + \frac{\partial G}{\partial \eta_s}\{\theta+(\eta_s-\theta+1)\ln(\eta_s-\theta+1)\}$$

$$= \frac{2\theta(2-\theta)}{(1-\theta)^3} + \frac{\pi}{3}\rho^*[g_0^2(1^+;\eta) - g_0^2(1^+;\theta)] + \frac{2\pi\rho^*}{3kT}\sum_i \varepsilon_i z_i^2 e^{z_i} f_2(\eta;z_i) \tag{27}$$

respectively. When $\sigma^3$, $\varepsilon_i$, and $z_i$ are independent of temperature, the above equations will reduce to

$$\beta\rho\frac{\partial F}{\partial\eta}(\eta-\theta+1)\ln(\eta-\theta+1) + \frac{\partial G}{\partial\eta_s}(\eta_s-\theta+1)\ln(\eta_s-\theta+1) = 0 \qquad (28)$$

and

$$\beta\rho\left[F+\theta\frac{\partial F}{\partial\eta}\right] + \theta\frac{\partial G}{\partial\eta_s} = \frac{2\theta(2-\theta)}{(1-\theta)^3} + \frac{\pi}{3}\overset{\bullet}{\rho}\left[g_0^2(1^+;\eta) - g_0^2(1^+;\theta)\right]$$
$$+ \frac{2\pi\overset{\bullet}{\rho}}{3kT}\sum_i \varepsilon_i z_i^2 e^{z_i} f_2(\eta;z_i). \qquad (29)$$

In fact, $z_i$ is independent of temperature if induced moments are neglected, and $\sigma^3$ and $\varepsilon_i$ usually are weak functions of temperature. Equations (26) and (27), or (28)-(29), consist of a set of nonlinear equations which are used to determine the reduced hard particle reduced densities $\eta$ and $\eta_s$. Having determined the values of $\eta$ and $\eta_s$, we can calculate the residual internal energy, compressibility factor, residual Helmholtz free energy, and the contact value of the RDF from Eqs. (10), (11), (20), and (7). These self-consistent values of $\eta$ and $\eta_s$ as functions of reduced density, $\rho^*$, for different isotherms, $(\beta\varepsilon_d)^{-1}$ or $T_r$, have similar results as those plotted in Figs.1 and 2 by Mansoori and Kioussis (1985).

## APPLICATIONS OF THE MODEL

For a simple non-polar fluid, like argon, there are no electrostatic interactions between the molecules. We thus use one simple Yukawa tail to represent the intermolecular dispersion potential. Note that for polar fluids we must include more Yukawa tails to represent the contributions due to electrostatic attractive intermolecular potential energies. The parameters of these Yukawas can be calculated directly from Eqs. (17) and (18). The resulting multiple-Yukawa model does not introduce any further computational difficulty.

We have chosen z=1.8 in parts of our calculations. As was reported by Henderson et al. (1978), the Yukawa fluid with this value is qualitatively similar to argon for the densities and temperatures of the vapor-liquid equilibrium. In addition, Stockmayer and Beattie (1942) have pointed out that the values of $B^*$ for the Lennard-Jones (12-6) potential in the range $1 < T^* < 5$ can be fitted with good accuracy by the equation

$$B^* T^{*\frac{1}{4}} = 1.064 - 3.602(1/T^*) \tag{30}$$

The above equation suggests that z is close to 1.8 when compared with Eq. (13).

Five different methods to determine the Yukawa potential parameters to represent the dispersion potential have been investigated: 1) fitting potential parameters $\varepsilon/k$ and $\sigma$ to the second virial coefficients when z=1.8 is fixed; 2) same as method1 but with the simple temperature dependence for the hard-core diameter; i.e.,

$$\sigma^3 = \sigma_0^3 \left(1 + \alpha/T^{*n}\right) \tag{31}$$

where $\sigma_0$ is a temperature independent hard-core diameter. Two more adjustable parameters $\alpha$ and n in addition to $\varepsilon/k$ and $\sigma_0$ are added in this case; 3) fitting potential parameters $\varepsilon/k$ and $\sigma$ to the second virial coefficients and z to $P_c = P_{eq}(T_c, \rho_c)$ (see Eq. (11)); 4) fitting potential parameters $\varepsilon/k$ and $\sigma$ to saturated liquid densities and vapor pressures with z=1.8;5) same as method 4 but including simple temperature dependence for the hard-core diameter, Eq. (31), and the value of n = 1.17is predetermined from fitting the second virial coefficients.

The characteristic parameters of Yukawa potential determined from fitting different fluid properties for argon are listed in Table I. In method 1, only temperatures above 133.15 K are used because the second virial coefficient has simple linear inverse temperature term in Eq. (13), which is too simple to represent low temperature second virial coefficient data. The experimental argon second virial coefficient data were taken from Michels et al. (1958) and Whalley et al. (1953). We have included low temperature second virial coefficient data by Fender and Halsey (1962) in method 2. Deviations between experimental and calculated second virial coefficients from method 2 over wide temperature range (84.79 - 873.15 K) are given in Table II. It can be seen that the calculated results are within the estimated experimental errors by including temperature dependence in the hard-core diameter. In method 3, we allow the range parameter z to float and a smaller $\varepsilon/k$ was found. The parameters in methods 4 and 5 were evaluated from fitting experimental argon vapor pressure data of Clark et al. (1951) and Michels et al. (1958) and saturated liquid density data of Michels et al. (1958) and Terry et al. (1969).  We have assigned the same weights for these two different types of data but weighted each data point differently by assessment of the quality of original data source.   We have also verified that the range parameter z is very close to 1.8 if it is allowed to be adjusted in methods4 and 5.   Deviations between experimental and calculated argon vapor pressures from method 4 are given in Table III and those of saturated liquid densities in Table IV.   In addition, we have included a molecular theory by Moser et al. (1981) for comparison in Table III.   As they are shown, predictions by this model are quite satisfactory.   The average percentage deviation of the vapor pressures is less than 3.76 % and that of saturatedl iquid densities is 2.17 %.   Also, the very small value of $\alpha$ in method V indicates the insignificance of including the temperature dependence of the hard-core diameter in fitting saturated properties.   Predictions of the compressibility factor of the vapor phase are generally in very good agreement with the experimantal data.   Such results obtained from the potential parameters of method 4 are the best; however, method 1 is simplest and can predict reasonable results.   Thus, this method is used to determine the Yukawa parameters for polar fluids as a prerequisite test of the extension of this model.

We have also applied this model to carbon dioxide ($CO_2$) and methyl chloride ($CH_3Cl$) with the Yukawa parameters determined by method 1.   Quadrupole moment value for $CO_2$, which is required to determine the second Yukawa parameters to represent the angle-averaged quadrupole-quadrupole interaction, was obtained from Hanna (1968). Since the quadrupole moment data are rather scarce and varying according to various sources, we have investigated the sensitivity of this value to the calculated results from the model.   For methylchloride, the polarizability and dipole moment data were taken from Moelwyn-Hughes (1961).

The predicted results in applying this model to carbon dioxide are accurate in the gas-like region.   Experimental second virial coefficients of carbon dioxide above 280 K given by Angus et al. (1973) were used to determine the Yukawa parameters to represent the dispersion potential.   Again, if $z_d$=1.8 is fixed, the fitted parameters are very close to those of the Lennard-Jones potential (Tee et al., 1966), of which similar results are obtained from argon.   These two Yukawa potential parameters are given in Table V.   It can be seen that $\varepsilon_1$ /k is dependent on temperature.   We have chosen an empirical Redlich-Kwong (RK) equation for comparison here.   The percentage deviations of the predicted pressures of the proposed model and those of RK equation of state for carbon dioxide at various temperatures are reported in Table VI. As it is shown, the proposed model give better results in the gas-like region. We note that variations in Q, which affect the values of $\varepsilon_1$ and $z_1$ in our formulations, Eqs. (17)-(18), only alter the calculated results slightly. The reason for the insensitivity of Q value is that only quadrupole-quadrupole interactions exist in carbon dioxide.

Next, experimental second virial coefficients of methyl chloride above 323.15 K (Sutter and Cole, 1970; Suh and Storvick, 1967) were used to determine the Yukawa parameters for dispersion potential, which are also given in Table V. A comparison of the percentage deviations between experimental pressures of $CH_3Cl$ (Hsu and Mcketta, 1964) and pressures calculated from this model or from RK equation of state for three isotherms are given in Table VII. These results indicate good agreement between experiment and the predictions from this model, given only the Yukawa intermolecular potential parameters and multipole moments. In the very high density region, this model is affected by the accuracy of the analytic form for the hard-sphere reference system and radial distribution function. Note that the Carnahan and Starling expression for the hard-sphere equation of state corresponds to a much softer fluid at high densities than is actually predicted from molecular dynamics. Better agreement with experimental results at high density region is expected if new and improved expressions for the hard-sphere reference system and radial distribution function are used. Also the assumption $z_d = 1.8$ for polar fluids needs to be verified.

**CONCLUSION**

The analytic statistical mechanical model for the hard-core with Yukawa tails has been shown to be successful for prediction of saturated properties of argon and most gas-like homogeneous region of non-polar and polar fluids. We have formulated attractive Yukawas to represent the electrostatic potentials in simple closed forms and applied these formulations to carbon dioxide and methylchloride. The present study indicates this model could be applied in prediction of different types of thermodynamic properties in gas-like region of polar fluids applicable for engineering design calculations.

**ACKNOWLEDGEMENT**

This research is supported by the Gas Research Institute Contract#5086-260-1244.

**NOMENCLATURE**

$\Delta a$  residual Helmholtz free energy per particle

$A^r$  residual Helmholtz free energy

$B$  second virial coefficient

$C_i$  constant in Eq. (18)

$d$  hard-sphere diameter

$g$  radial distribution function

$g$(With Hat) Laplace transform

$k$  Boltzmann's constant

$n$  adjustable parameter in Eq. (31)

$P$  pressure

$Q$  quadrupole moment

$r$  intermolecular separation distance

$r_n$  maximum intermolecular separation distance in our                    calculating procedure of $\alpha_n/r^n$ term

$R$  universal gas constant

$\Delta s$  residual entropy per particle

$T$  thermodynamic temperature

$T_r$  reduced temperature, $=T/T_c$

$\Delta u$  residual internal energy per particle

$U^r$  residual internal energy

$\bar{u}$  angle-averaged energy function

$x$  dimensionless quantity, $= r/\sigma$

$z$  inverse range parameter

**GREEK LETTERS**

$\alpha$  adjustable parameter in Eq. (31)

$\alpha_i$  polarizability of the molecule i

$\alpha_6$  angle-averaged electrostatic attractive potential of                    separation distance $r^{-6}$ term

$\beta$  $=1/kT$

$\varepsilon$  parameter describing the depth of attractive well in the intermolecular interaction

$\mu$  dipole moment

$\rho$  number density

$\sigma$  diameter of the hard core of the molecules

$\sigma_0$  temperature independent hard-core diameter

$\theta$  packing fraction

$\eta$  reduced density, $=\pi\rho d^3/6$

$\phi$  intermolecular potential

**SUBSCRIPTS**

0  hard-sphere quantities
c  critical properties
d  dispersion potential
i  molecule i or Yukawa potential i
LJ Lennard-Jones potential
s  hard-particle properties related to the residual entropy

**SUPERSCRIPTS**

\*  dimensionless quantities
r  residual properties

**Table I.** Characteristic parameters of Yukawa potential determined from fitting different fluid properties for argon

| Method | $\sigma$ or $\sigma_0$ $A$ | $\varepsilon/k$ $K$ | $z$ | $\alpha$ |
|--------|------|--------|-------|-----------------|
| 1 | 3.353 | 116.72 | 1.8 | |
| 2 | 3.033 | 113.99 | 1.8 | 0.594 |
| 3 | 3.353 | 54.58 | 0.931 | |
| 4 | 3.474 | 122.98 | 1.8 | |
| 5 | 3.471 | 122.92 | 1.8 | $3.75 \times 10^{-7}$ |

Table II.  Deviations between experimental and calculated argon second
virial coefficients from method 2 in Table I

| T<br>K | B<br>cm³/mol | (B−B$_{cal}$/B)×100 | T<br>K | B<br>cm³/mol | (B−B$_{cal}$/B)×100 |
|---|---|---|---|---|---|
| 84.79 | −249.34 | 0.54 | 153.15 | −82.97 | −0.86 |
| 88.34 | −229.89 | −0.21 | 163.15 | −73.25 | −0.88 |
| 92.30 | −211.79 | −0.54 | 173.15 | −65.21 | −0.51 |
| 95.06 | −200.87 | −0.54 | 188.15 | −54.83 | −0.47 |
| 101.40 | −178.73 | −0.59 | 203.15 | −46.52 | −0.07 |
| 102.01 | −177.65 | −0.12 | 223.15 | −37.43 | 0.22 |
| 105.51 | −166.06 | −0.77 | 248.15 | −28.57 | 0.75 |
| 108.15 | −160.27 | 0.15 | 273.15 | −22.41 | 4.85 |
| 113.32 | −149.58 | 1.72 | 323.15 | −11.20 | 1.27 |
| 117.50 | −140.58 | 2.11 | 373.15 | −4.34 | 9.22 |
| 123.99 | −127.99 | 2.59 | 423.15 | 1.01 | −26.49 |
| 133.15 | −107.98 | −1.15 | 473.15 | 5.28 | 0.32 |
| 138.15 | −100.88 | −1.03 | 573.15 | 10.77 | −1.64 |
| 143.15 | −94.42 | −0.92 | 673.15 | 15.74 | 5.94 |
| 148.15 | −88.45 | −0.88 | 773.15 | 17.76 | 0.92 |
| 150.65 | −85.63 | −0.89 | 873.15 | 19.48 | −1.17 |

**Table III.** **Deviations between experimental and calculated argon vapor pressures from method 4 in Table I**

| T<br>K | P<br>bar | $P - P_{cal}$<br>bar | $(P - P_{cal}/P) \times 100$ |
|---|---|---|---|
| 132.759 | 23.076 | - 0.989 | - 4.288 |
| 127.715 | 18.162 | - 0.776 | - 4.273 |
| 125.182 | 16.000 | - 0.657 | - 4.104 |
| 120.035 | 12.180 | - 0.418 | - 3.431 |
| 115.754 | 9.540 | - 0.240 | - 2.514 |
| 109.637 | 6.515 | - 0.037 | - 0.570 |
| 107.529 | 5.656 | 0.017 | 0.303 |
| 101.491 | 3.648 | 0.121 | 3.330 |
| 98.349 | 2.843 | 0.153 | 5.383 |
| 93.178 | 1.815 | 0.171 | 9.413 |

**Table IV.** Deviations between experimental and calculated argon saturated liquid densities from method 4 in Table I

| T<br>K | $\rho$<br>mol/cm³ | $\rho - \rho_{cal}$<br>mol/cm³ | $(\rho - \rho_{cal}/\rho) \times 100$ |
|---|---|---|---|
| 134.760 | 0.025352 | 0.001163 | 4.589 |
| 127.480 | 0.027344 | 0.001100 | 4.024 |
| 125.051 | 0.027905 | 0.001014 | 3.632 |
| 119.890 | 0.029097 | 0.000878 | 3.016 |
| 116.821 | 0.029700 | 0.000716 | 2.411 |
| 109.408 | 0.031126 | 0.000356 | 1.143 |
| 105.420 | 0.031905 | 0.000200 | 0.628 |
| 100.660 | 0.032727 | - 0.000076 | - 0.231 |
| 97.452 | 0.033213 | - 0.000322 | - 0.968 |
| 95.790 | 0.033542 | - 0.000369 | - 1.100 |
| 92.148 | 0.034068 | - 0.000666 | - 1.955 |
| 89.919 | 0.034416 | - 0.000820 | - 2.383 |

Table V. Potential parameters for Ar, $CO_2$ and $CH_3Cl$ determined from second virial coefficients

| Molecular pair | Yukawa potential | | | | | L-J potential | |
|---|---|---|---|---|---|---|---|
| | $\sigma$ A | $\epsilon_1/k$ K | $z_1$ | $\epsilon_2/k$ K | $z_2$ | $\sigma$ A | $\epsilon/k$ K |
| Ar | 3.353 | 116.72 | 1.8 | | | 3.499 | 118.13[*] |
| $CO_2$ | 3.730 | 170.90 | 1.8 | 70699.5/T | 7.29 | 4.416 | 192.25[*] |
| $CH_3Cl$ | 5.734 | 195.23 | 1.8 | 13885.86/T+ 14.9965 | 3.05 | 4.449 | 228.2[ζ] |

[*] reported by Tee et al. (1966).
[ζ] reported by Moelwyn-Hughes (1961).

**Table VI.** Predicted pressures of carbon dioxide from the model and experimental data reported by Angus et al. (1976)

| T K | V cm³/mol | $P_{exp}$ bar | $P_{cal}$ bar |
|---|---|---|---|
| 230 | 18894 | 1.0 | 1.0033 |
| 270 | 11067 | 2.0 | 2.0073 |
| 300 | 4863.9 | 5.0 | 5.0344 |
| 350 | 2823.4 | 10 | 13.095 |
| 400 | 1601.9 | 20 | 20.284 |
| 400 | 770.3 | 40 | 41.178 |
| 450 | 581.87 | 60 | 62.121 |
| 500 | 492.16 | 80 | 83.147 |
| 500 | 389.14 | 100 | 105.01 |
| 500 | 253.28 | 150 | 161.74 |

**Table VII.** Predicted pressures of methyl chloride from the model and experimental data reported by Hsu and Mcketta (1964)

| T K | V cm³/mol | P_exp bar | P_cal bar |
|---|---|---|---|
| 323.15 | 3999.9 | 6.142 | 6.3028 |
|  | 2875.5 | 8.249 | 8.5486 |
|  | 2589.0 | 9.081 | 9.4001 |
|  | 2245.0 | 10.372 | 10.674 |
| 348.15 | 4331.6 | 6.255 | 6.3652 |
|  | 2525.6 | 10.218 | 10.544 |
|  | 1533.2 | 15.551 | 16.460 |
|  | 1139.6 | 19.445 | 21.138 |
| 373.15 | 4733.1 | 6.261 | 6.318 |
|  | 2703.2 | 10.536 | 10.766 |
|  | 1639.9 | 16.345 | 17.048 |
|  | 1257.5 | 20.330 | 21.577 |